\documentclass[showpacs,onecolumn]{revtex4}
\usepackage{graphicx}
\usepackage{subfigure}
\usepackage{caption2}
\usepackage{float}
\usepackage{amsmath}
\usepackage{amssymb}

\begin{document}

\title{Stabilization of two-dimensional solitons and vortices against
supercritical collapse by lattice potentials}
\author{Rodislav Driben$^{1}$ and Boris A. Malomed$^{2}$}
\affiliation{Laboratoire de Photonique Quantique et Mol\'{e}culaire, CNRS, \'{E}cole
Normale Sup\'{e}rieure de Cachan, UMR 8537, 94235 Cachan, France}
\affiliation{$^{2}$Department of Physical Electronics, School of Electrical Engineering,
Faculty of Engineering, Tel Aviv 69978, Israel}

\begin{abstract}
It is known that optical-lattice (OL) potentials can stabilize
solitons and solitary vortices against the critical collapse,
generated by the cubic attractive nonlinearity in the 2D geometry.
We demonstrate that OLs can also stabilize various species of
fundamental and vortical solitons against the \emph{supercritical}
collapse, driven by the double-attractive cubic-quintic
nonlinearity (however, solitons remain unstable in the case of the
pure quintic nonlinearity). Two types of OLs are considered,
producing similar results: the 2D Kronig-Penney ``checkerboard",
and the sinusoidal potential. Soliton families are obtained by
means of a variational approximation, and as numerical solutions.
The stability of the families, which include fundamental and
multi-humped solitons, vortices of
oblique and straight types, vortices built of quadrupoles, and \textit{%
supervortices}, strictly obeys the Vakhitov-Kolokolov criterion. The
model applies to optical media and BEC in ``pancake" traps.
\end{abstract}

\maketitle

\section{Introduction and the model\textit{\ }}

Formation of multidimensional solitons and solitary vortices (solitons with
embedded vorticity) has drawn a great deal of attention in studies of
nonlinear optics and Bose-Einstein condensation (BEC), see review \cite%
{review}. (2+1)-dimensional spatial solitons and quasi-2D spatiotemporal
ones were created in crystals with the quadratic ($\chi ^{(2)}$)
nonlinearity \cite{chi2spatiotemp}. Also reported were spatial solitons,
vortices, and dipole-mode states in photorefractive crystals with a
photo-induced lattice, where the nonlinearity is saturable \cite{photorefr}.
However, truly 2D or 3D solitons have not yet been observed in media with
the generic cubic ($\chi ^{(3)}$) nonlinearity, the problem being their
instability against the collapse \cite{Talanov,collapse}. Vortex solitons
are subject (in the uniform space) to a still stronger azimuthal
instability, which occurs even in the absence of the collapse \cite{Skryabin}%
.

A relevant ingredient of both optical and BEC settings which may stabilize
multidimensional solitons and vortices is an effective periodic potential.
In optics, it implies a periodic modulation of the refractive index in the
transverse plane, as in photonic-crystal fibers), and in BEC it is induced
by optical lattices (OLs). In the medium with the self-focusing (SF) $\chi
^{(3)}$ nonlinearity, the stabilization of fundamental and vortical 2D
solitons (those with topological charges $S=0$ and $1$, respectively) under
the action of the square-lattice potential was predicted in works \cite{BBB1}
and \cite{Jianke} (see also works \cite{Barcelona}; stable solitons were
found too in photonic-crystal-fiber models \cite{PhotCryst}). The action of
the stabilization mechanism can be summarized as follows. In the free 2D
space, the $\chi ^{(3)}$ nonlinearity supports a family of Townes solitons
(TSs), which have the same norm, $Q=Q_{\mathrm{TS}}$, at all values of the
propagation constants, $k$ \cite{Townes,collapse}. Although there are no
unstable eigenvalues in the spectrum of small perturbations around Townes
solitons, a specific zero eigenvalue accounts for the instability against
sub-exponentially growing perturbations, which is a manifestation of the
\emph{critical character} of the collapse induced by the $\chi ^{(3)}$
nonlinearity in the 2D space. As demonstrated by numerical findings and the
variational approximation (VA), the action of the OL potential with small
strength $\epsilon $ lifts the degeneracy of the Townes-soliton family,
stretching the single point, $Q=Q_{\mathrm{TS}}$, into interval $Q_{\mathrm{%
TS}}-\Delta Q<Q<Q_{\mathrm{TS}}$ of width $\Delta Q\sim \epsilon $, which is
filled by stable solitons, that obey the Vakhitov-Kolokolov (VK) criterion, $%
dQ/dk>0$. This criterion is well known to be a necessary condition for the
stability of solitons in media with SF nonlinearities \cite{VK,collapse}.

As concerns vortex solitons supported by the square OL, the most
compact ``crater-shaped" ones, represented by a single density peak
with a vorticity-induced hole in the center, are unstable
\cite{Cachan}. Stable vortices with $S=1$ can be built as sets of
four (or eight) peaks, with the phase shift $\pi /2$ (or $\pi /4$,
respectively) between adjacent ones \cite{BBB1,Jianke}. Stable
vortices of higher orders, up to $S=6$, built of up to $12$ peaks,
were found too, as well as stable ``supervortices", i.e., ring
chains of $12$ (or more) compacts vortices carrying local spins
$s=1$, with global vorticity $S=\pm 1$ imprinted onto the entire
ring. These patterns were found in models with cubic and saturable
SF nonlinearities \cite{HS}. Stable vortices of the gap-soliton type
were also found in the model combining the OL and self-repulsive
nonlinearity \cite{gap-vortex}. A general mathematical
classification of various localized states with an intrinsic phase
structure (in particular, vortices) in the 2D model with the
square-lattice $\cos $ potential and cubic nonlinearity was
developed in \cite{Jianke2}.

In the 2D geometry, \textit{supercritical collapse} is generated by the SF
quintic ($\chi ^{(5)}$) nonlinearity, which may occur in a combination with $%
\chi ^{(3)}$ terms. In optics, the cubic-quintic (CQ) nonlinearity with the
SF $\chi ^{(5)}$ part was predicted \cite{Agarwal} and recently observed
\cite{Brazil} in aqueous colloids. It was also observed in dye solutions
\cite{dye}, and recently predicted, through the cascaded mechanism, in
two-level media \cite{Rochester}. It is relevant to mention that the $\chi
^{(5)}$ nonlinearity with the self-defocusing sign is observed in various
uniform optical media \cite{opticalCQ}.

Besides the context of nonlinear optics, the CQ nonlinearity of the SF type
appears in the description of BEC with attraction between atoms trapped in a
``pancake" configuration. Strictly speaking, the reduction of the underlying
3D Gross-Pitaevskii equation (GPE) to the 2D form produces a more complex
nonpolynomial nonlinearity \cite{Luca}. However, for 1D configurations
corresponding to a cigar-shaped trap, it was demonstrated that the
respective GPE with the SF quintic term, which represents a ``vestige" of
the underlying multi-dimensionality, provides for an appropriate description
of the soliton dynamics, unless one is interested in the collapse per se
\cite{BEC-CQ}. Another straightforward interpretation of the quintic term in
the GPE is the contribution of three-body collisions, provided that the
lossy part of this interaction (kicking out atoms from the condensate) may
be neglected \cite{three-body}.

The OL potential is not necessary for the stability of 1D solitons in the SF
CQ model, as they are stable in the free 1D space, despite the possibility
of the collapse \cite{Seva}. On the contrary, in the 2D model with the SF $%
\chi ^{(3)}$ and $\chi ^{(5)}$ nonlinearities, the OL is a crucial factor
for the stabilization of solitons against the supercritical collapse. In the
normalized form, the corresponding model is based on the equation for the
mean-field wave function, $u(x,y,t)$ (or the local amplitude of an
electromagnetic wave, in terms of optics):

\begin{equation}
iu_{t}+u_{xx}+u_{yy}-V(x,y)u+2|u|^{2}u+\gamma |u|^{4}u=0.  \label{NLS}
\end{equation}%
The derivation of Eq. (\ref{NLS}) from the full GPE yields $\gamma \sim $ $%
a_{z}^{2}/D^{2}$, where $a_{z}$ is the width of the transverse confinement,
and $D$ the half-period of the OL\ potential. We will consider two different
potentials,\textit{\ viz}., the cosinusoidal ($\cos $) one, $V=-V_{0}\left[
\cos \left( \pi x/D\right) +\cos \left( \pi y/D\right) \right] $, and the 2D
Kronig-Penney (KP) potential, in the form of a ``checkerboard" composed of
cells of size $D$ with potential difference $V_{0}$ between adjacent ones
\cite{Cachan} (2D states in the checkerboard potential combined with the CQ
nonlinearity in which the $\chi ^{(5)}$ term is self-defocusing, i.e., $%
\gamma <0$, hence the collapse does not occur, were recently investigated in
Ref. \cite{Cachan}). Equation (\ref{NLS}) conserves the norm,
\begin{equation}
Q\equiv \int \int |u(x,y)|^{2}dxdy.  \label{Q}
\end{equation}%
Undoing scalings used in the derivation of Eq. (\ref{NLS}) from the
underlying GPE, one concludes that the actual number of atoms in the BEC is $%
N\simeq \left( a_{z}/\left\vert a_{s}\right\vert \right) Q$, where $a_{s}$
is the scattering length. In terms of nonlinear optics, evolution variable $%
t $ in Eq. (\ref{NLS}) is the propagation distance, rather than time.

For the purpose of comparison, we will also consider a modification of Eq. (%
\ref{NLS}) with pure quintic nonlinearity,

\begin{equation}
iu_{t}+u_{xx}+u_{yy}-V(x,y)u+|u|^{4}u=0.  \label{quintic}
\end{equation}%
Stationary solutions to Eqs. (\ref{NLS}) and (\ref{quintic}) for 2D
fundamental (single-humped) solitons were found by means of the VA and in a
numerical form, as reported below in Section II. Higher-order (multi-humped)
nontopological solitons, as well as vortices (including those built of
quadrupoles, rather than of fundamental solitons) and supervortices, were
constructed by means of numerical methods, see Sections II and III,
respectively. The stability of all these localized patterns was inferred
from the VK criterion and verified by direct simulations. To test the
stability numerically, the initial perturbation was, typically, imposed by
the multiplication of a numerically exact stationary solution by perturbing
factor $\left( 1+\varepsilon \right) $, typically with $\left\vert
\varepsilon \right\vert \lesssim 0.03$ (this was quite sufficient to
identify stable and unstable solitons, monitoring their evolution in the
course of sufficiently long simulations). As a result, a border between
stable and unstable solutions has been found in each soliton family in the
CQ model, while all solitons of Eq. (\ref{quintic}) are unstable.
Approaching the stability border, the width of the \textit{stability margin}%
, i.e., the maximum value of perturbation amplitude $\varepsilon $ in the
above expression, which does not trigger the onset of instability, shrinks
to zero.

\section{Fundamental and multi-humped solitons}

\begin{figure}[tbp]
\begin{center}
{\includegraphics[width=4.5in]{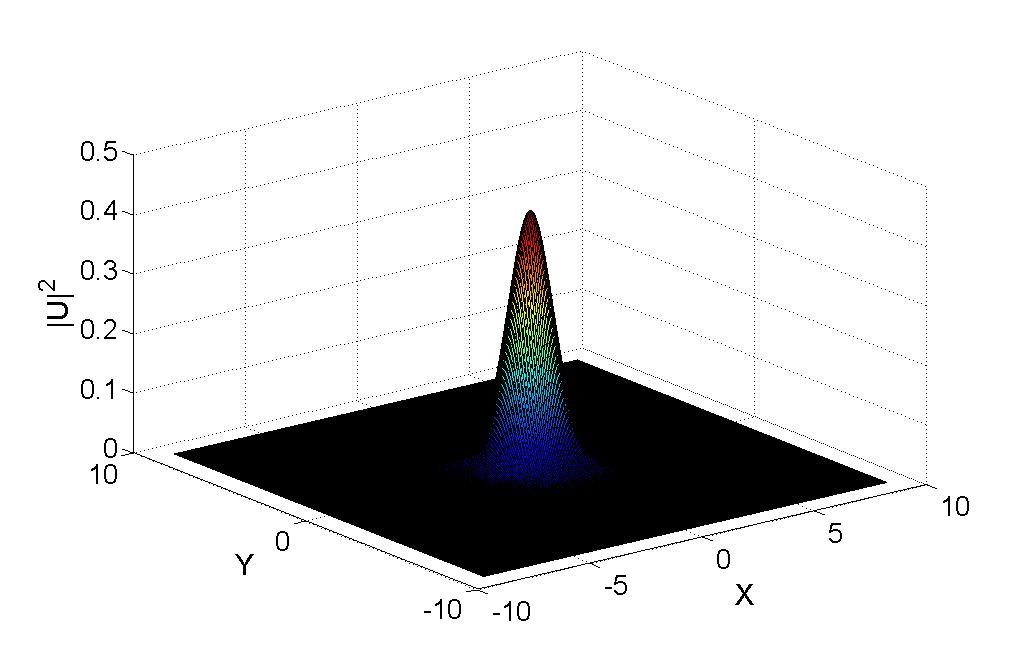}}
\end{center}
\caption{(Color online) A typical example of numerically found stable
fundamental soliton in the model with the KP (checkerboard) potential, for $%
D=3$, $V_{0}=2$, and $\protect\gamma =1$. The soliton is pertains to $k=1.53$%
, its norm being $Q=2.05.$}
\label{fig1}
\end{figure}

Solutions to Eq. (\ref{NLS}) with real chemical potential $-k$ (in optical
models, $k$ is the propagation constant) are sought for as $u(x,z)=\exp
\left( ikt\right) U(x,y)$, with $U(x,y)$ obeying the stationary equation,
\begin{equation}
U_{xx}+U_{yy}-V(x,y)U+2|U|^{2}U+\gamma |U|^{4}U=kU,  \label{stat}
\end{equation}%
which is associated with Lagrangian $L=\int \int L(x,y)dxdy$, whose density
is%
\begin{equation}
\mathit{L}=\left\vert U_{x}\right\vert ^{2}+\left\vert U_{y}\right\vert ^{2}+%
\left[ k+V\left( x,y\right) \right] \left\vert U\right\vert ^{2}-\left\vert
U\right\vert ^{4}-\left( \gamma /3\right) \left\vert U\right\vert ^{6}.
\label{L}
\end{equation}%
Fundamental solitons with amplitude $A$ and width $W$ may be approximated by
ansatz $U\left( x,y\right) =A\exp \left( -\left( x^{2}+y^{2}\right) /\left(
2W^{2}\right) \right) $, whose norm (\ref{Q}) is $Q=\pi A^{2}W^{2}$, hence
the ansatz can be written as%
\begin{equation}
U(x,y)=\frac{\sqrt{Q/\pi }}{W}\exp \left( -\frac{x^{2}+y^{2}}{2W^{2}}\right)
~,  \label{ans}
\end{equation}%
where $Q$ and $W$ may be treated as free variational parameters. The use of
the isotropic ansatz is suggested by numerically found shapes of the
fundamental solitons, which feature an approximate axial symmetry, except
for ``tails", where the local amplitude of the wave field is very small, see
Fig. \ref{fig1}.

Subsequent derivation of the VA from Lagrangian (\ref{L}) and ansatz (\ref%
{ans}) is straightforward for the $\cos $ potential. In the KP model, the
checkerboard potential may be replaced, for this purpose, by its two lowest
spatial harmonics \cite{Cachan}. Thus, the calculation of the respective
effective Lagrangian, $L=L\left( Q,W\right) $, leads to the variational
equations, $\partial L/\partial Q=\partial L/\partial W=0$, which take the
form of%
\begin{eqnarray}
k+\frac{1}{W^{2}}-\frac{V_{0}}{2}\left[ 1+\exp \left( -\frac{\pi ^{2}W^{2}}{%
4D^{2}}\right) \right] -\frac{Q}{\pi W^{2}}-\frac{\gamma Q^{2}}{3\pi
^{2}W^{4}} &=&0,  \notag \\
&&  \label{vareqns} \\
1-\frac{Q}{2\pi }-\frac{2\gamma Q^{2}}{9\pi ^{2}W^{2}}-\frac{\pi
^{2}V_{0}W^{4}}{8D^{2}}\exp \left( -\frac{\pi ^{2}W^{2}}{4D^{2}}\right) &=&0.
\notag
\end{eqnarray}%
The modification of the VA for the quintic-only equation (\ref{quintic}) is
obvious: in Eqs. (\ref{vareqns}), the terms linear in $Q$ should be dropped,
and $\gamma =1$ should be substituted.

\begin{figure}[tbp]
\begin{center}
\includegraphics[width=4.5in]{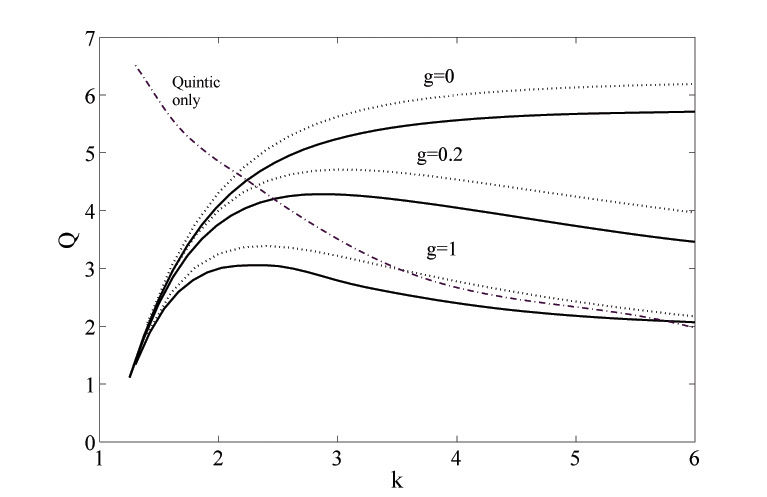}
\end{center}
\caption{The norm of the fundamental solitons versus the propagation
constant, $k$, as predicted by the VA (dotted curves) and found from
numerical solution of Eq. (\protect\ref{stat}) with the $\cos $ potential
(solid curves). Fixed parameters are $V_{0}=2$ and $D=3$. In addition, the
dashed-dotted curve displays the same dependence as predicted by the VA for
the model with the pure quintic nonlinearity, i.e., Eq. (\protect\ref%
{quintic}).}
\label{fig2}
\end{figure}

Families of fundamental solitons, as predicted by the VA, i.e., obtained
from a numerical solution of Eqs. (\ref{vareqns}), and found from a
numerical solution of Eq. (\ref{stat}), that was performed by means of a
modification of the relaxation method, are represented in Fig. \ref{fig2} by
a set of curves $Q(k)$, for different values of $\chi ^{(5)}$ coefficient $%
\gamma $. The case of the pure quintic nonlinearity, which corresponds to
Eq. (\ref{quintic}), is included too.

It is seen that solitons in the CQ model do not exist with the norm below a
threshold value, $Q_{\min }$, which is explained by the delocalization
transition \cite{Salerno} in the region dominated by the cubic nonlinearity
(therefore $Q_{\min }$ weakly depends on $\gamma $). Further, the VK
criterion suggests that, in the presence of the cubic term and for any $%
\gamma >0$, the solitons are stable only up to a point at which $dQ/dk$
changes the signs. In accordance with works \cite{BBB1}, the
stability-change point does not exist in the model with $\gamma =0$. Note
that the norm of stable solitons attains the largest value, which is $Q=Q_{%
\mathrm{TS}}$, for $\gamma =0$; with the growth of $\gamma $, the largest
norm of the stable solitons decreases, which demonstrates the increasing
difficulty in the stabilization of the solitons with the transition from the
critical (cubic) to supercritical (quintic) nonlinearity in the 2D geometry.
In accordance with this trend, all solitons in the pure quintic model (the
one without the cubic term) are unstable (as per the VK criterion), as the $%
Q(k)$ curve for the quintic model (the dashed-dotted curve in Fig. \ref{fig2}%
) features solely the negative slope, $dQ/dk<0$. Direct simulations exactly
corroborate all the predictions of the VK criterion. While Fig. \ref{fig2}
displays the results for the model with the $\cos $ potential, the situation
in the KP model is the same, the respective $Q(k)$ curves being very close
to those shown in Fig. \ref{fig2}.

In the CQ\ model, the decrease of $V_{0}$ leads to shrinkage of the portions
of the $Q(k)$ curves with the positive slope, and they disappear at some $%
\left( V_{0}\right) _{\min }$, leaving only unstable states, with $%
dQ/dk<0\allowbreak $. The respective stability regions for the fundamental
solitons in the $\left( V_{0},Q\right) $ plane are displayed in Fig. \ref%
{fig3} for the KP model (the situation for its counterpart with the $\cos $
potential is very similar). The lower stability boundary in this figure is
the delocalization border, below which solitons do not exist, while the
upper border is exactly predicted by the VK criterion (i.e., solitons are
unstable above it). The evident trend to the shrinkage of the stability
region with the decrease of $D$ is explained by the exponential smallness of
the force of the interaction of a broad soliton with a short-period OL.
\begin{figure}[tbp]
\begin{center}
\includegraphics[width=4.5in]{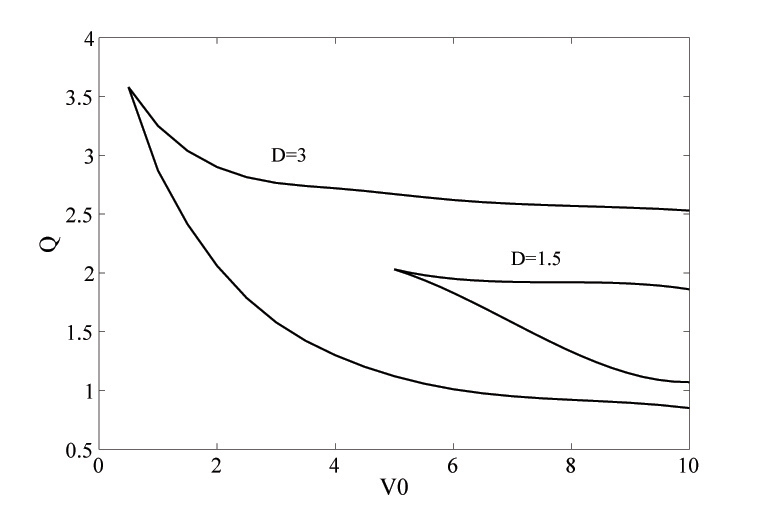}
\end{center}
\caption{Stability regions (between the upper and lower borders) for
fundamental solitons in the plane of the OL strength ($V_{0}$) and soliton's
norm ($Q$) in the KP model for different values of OL half-period $D$. The
coefficient in front of the quintic term is $\protect\gamma =1$.}
\label{fig3}
\end{figure}

Values of $k$ for all solutions reported in this work belong to the
semi-infinite gap in the OL-induced spectrum. Solitons can also be found in
finite bandgaps; however, as well as in the 2D KP model with the
self-repulsive $\chi ^{(5)}$ term \cite{Cachan}, all gap solitons turn out
to be unstable.

\begin{figure}[tbp]
\begin{center}
\subfigure[]{\includegraphics[width=3.6in]{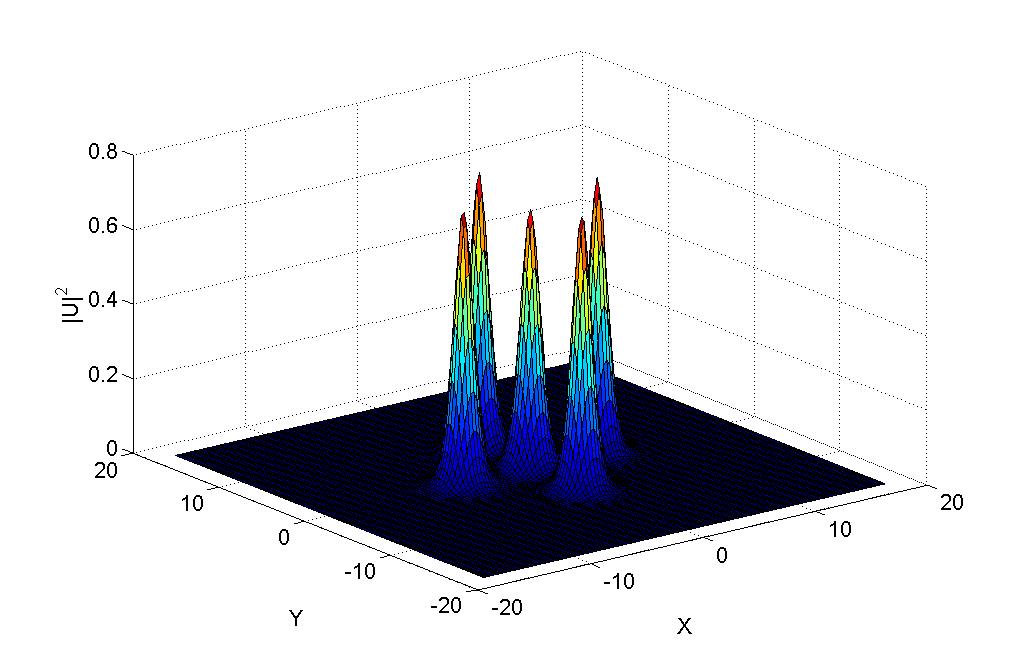}}\subfigure[]{%
\includegraphics[width=3.6in]{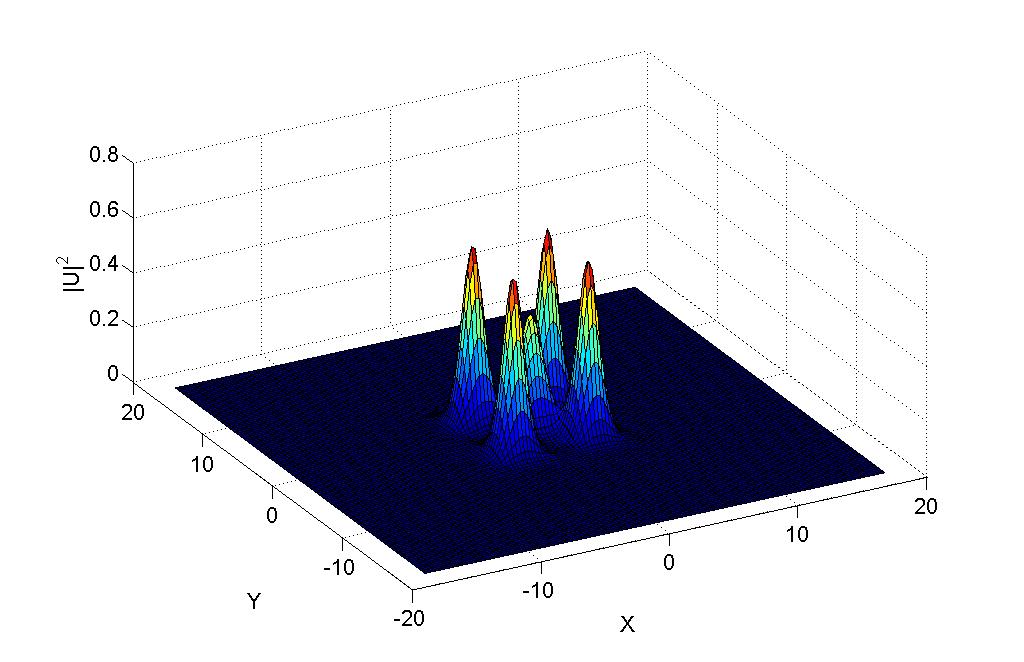}}
\end{center}
\caption{(Color online) Typical examples of the five-humped soliton found in
models with the KP (b) and cosinusoidal (b) potential, respectively. Both
solitons pertain to propagation constant $k=1.8$ and differ in the total
norm: $Q=12.42$ for panel (a), and $Q=14.44$ for (b). Lattice parameters are
the same as those corresponding to Fig. \protect\ref{fig1}.}
\label{fig4}
\end{figure}

As shown above, the supercritical collapse imposes an upper bound on the
norm of stable fundamental solitons [undoing rescalings leading to Eq. (\ref%
{NLS}), one can conclude that the number of atoms in the respective
matter-wave soliton is $\lesssim 10^{4}$]. Stable localized states with a
larger norm can be built as multi-humped solitons. Due to the symmetry
imposed by the OL in two dimensions, the first species of that type
following fundamental solitons features five peaks, cf. Ref. \cite{Cachan}.
Figure \ref{fig4} displays examples of stable five-peaked solitons found in
the KP and $\cos $ models, for a common value of the chemical potential, $%
k=1.8$ (the solitons are rotated relative to each other by angle $\pi /4$
due to the difference in the definition of the periodic potential in the
models). As well as the fundamental solitons, families of these solutions
feature the stability-change point, separating portions of the respective $%
Q(k)$ curves with $dQ/dk>0$ and $<0$ (not shown here). Actually, dependences
$Q(k)$ and stability regions in the plane of $\left( V_{0},Q\right) $ for
the five-peak solitons are similar to those for the fundamental solitons,
which are displayed above in Figs. \ref{fig2} and \ref{fig3}, with a
difference that $Q$ is larger by a factor of $\simeq 4$. Injection of more
norm gives rise to higher-order solitons. For $Q$ still larger than the
maximum value admitted by the five-peak solitons, their nine-peak
counterparts were obtained.

Besides the solutions built as complexes of in-phase peaks, the model also
gives rise to stable dipole, quadrupole, and multi-pole localized states. In
particular, examples of a quadrupole can be seen below in Fig. \ref{fig8},
as building blocks used to compose a new type of vortices (\textit{%
quadrupole vortices}).

\section{Vortex solitons, quadrupole vortices, and supervortices}

\begin{figure}[tbp]
\begin{center}
\includegraphics[width=4.5in]{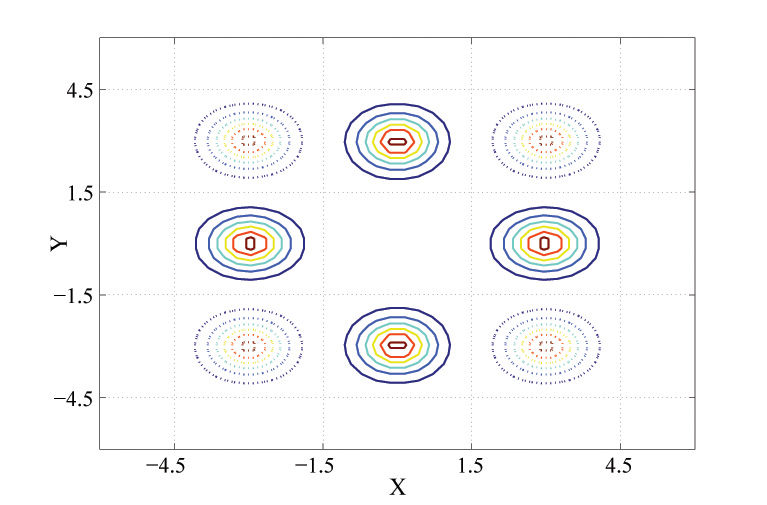}
\end{center}
\caption{(Color online) Juxtaposition of contour plots showing density
levels, $|U\left( x,y\right) |^{2}$, in straight and oblique vortices (solid
and dotted curves, severally). The vortices appertain to $V_{0}=10$ and $%
k=10 $, with norms $9.79$ and $9.67$ for the oblique and straight ones,
respectively. }
\label{fig5}
\end{figure}

Solitary vortices with topological charge $S$ are found as complex
solutions to Eq. (\ref{stat}) with the phase circulation of $2\pi
S$. Compact (``crater-shaped") vortex solitons, with the vorticity
nested in a single peak, are unstable (the ``crater" splits into a
set of nonsteady pulses resembling fundamental soltions, with a
single one surviving the subsequent evolution). Two species of
stable
vortices with $S=1$ have been found in the present models (with the KP and $%
\cos $ potentials alike), either one being arranged as a set of four
peaks with phase shifts $\pi /2$ between them. Referring to the
orientation of diagonals connecting the opposite peaks, which
compose the vortices, relative to the KP ``checkerboard", the
species may be called \textit{oblique} and \textit{straight},
examples of which are displayed together in Fig. \ref{fig5}. While
oblique vortices include a nearly empty site at the center, the
straight vortex places its center at a local potential maximum,
without any vacancy.

\begin{figure}[tbp]
\begin{center}
\includegraphics[width=4.5in]{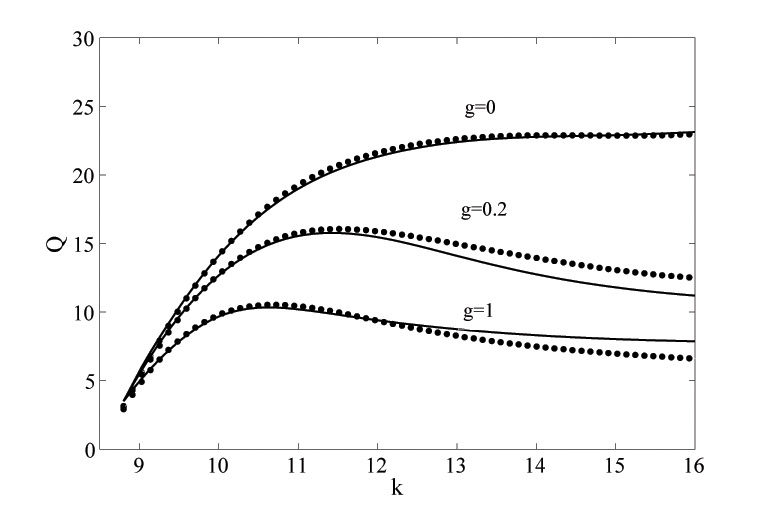}
\end{center}
\caption{$\ Q(k)$ curves for families of oblique and straight vortex
solitons (dotted and solid curves, respectively), with $S=1$, in the KP
model with $V_{0}=10,D=3,\protect\gamma =1$.}
\label{fig6}
\end{figure}

Families of the solitary vortices of both types are presented in Fig. \ref%
{fig6} by the respective $Q(k)$ curves, which summarize numerically obtained
results. The VK criterion is only a necessary condition for the stability of
vortices, as it does not detect azimuthal instabilities \cite{review}.
Nevertheless, systematic simulations have demonstrated that the stability of
the vortex solitons in the present model precisely obeys the VK criterion,
i.e., they are stable up to the point where $dQ/dk$ changes its sign, see
Fig. \ref{fig6}. This is explained by the fact that the lattice potential is
strong enough to suppress the azimuthal instability of the vortices, similar
to the situation in the cubic model \cite{BBB1}. Note that, as well as in
the case of fundamental solitons, cf. \ref{fig2}, the stability-change
point, $dQ/dk=0$, does not exist in the limit of the cubic equation, i.e., $%
\gamma=0$, and the norm of the solitary vortices attains its maximum just in
this limit.

\begin{figure}[tbp]
\begin{center}
\subfigure[]{\includegraphics[width=3.6in]{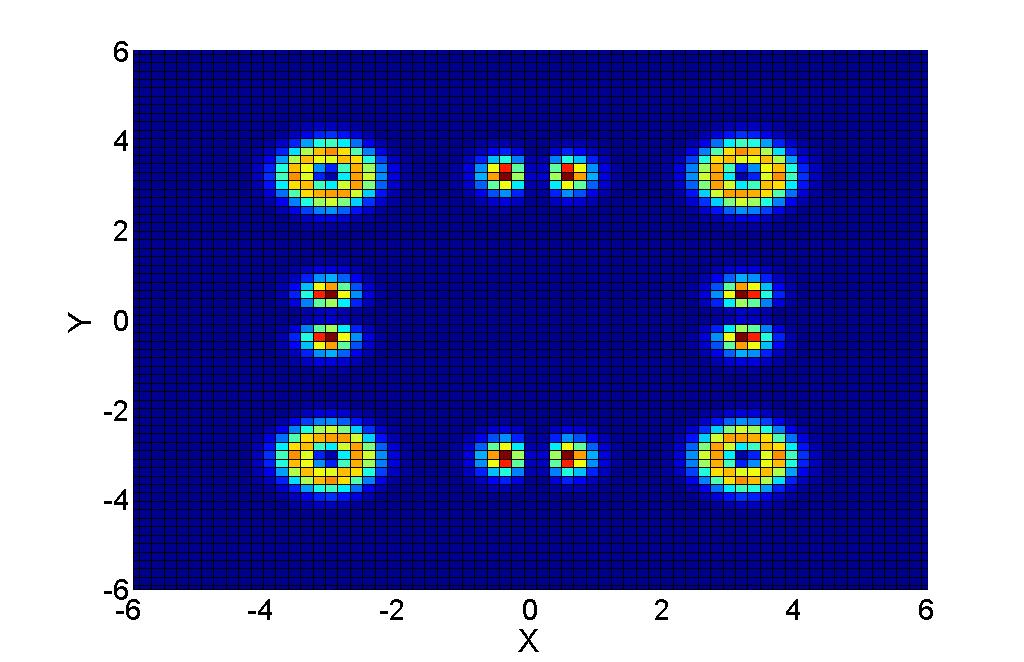}}\subfigure[]{%
\includegraphics[width=3.6in]{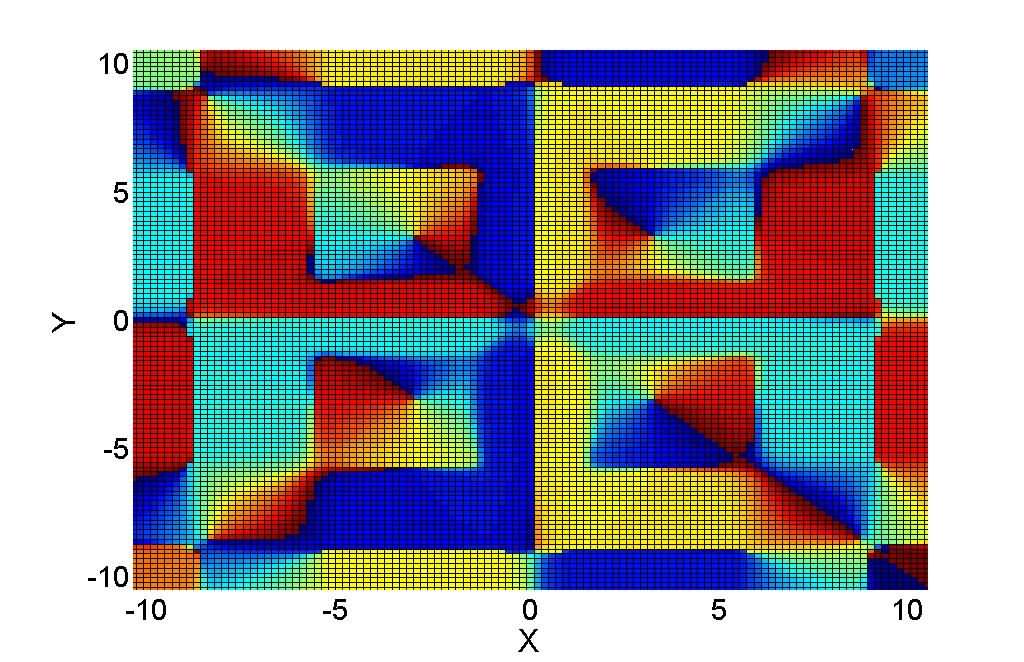}}
\end{center}
\caption{(Color online) \ The distribution of the local density (a) and
phase (b) in a stable supervortex supported by the KP potential with $%
V_{0}=10$, $D=2$, and $\protect\gamma =1$, for $k=9$ and $Q=24.26$. The
global vorticity of the pattern is $S=+1$, while spins of four individual
crater-shaped vortices, with centers placed at points $\left( x,y\right)
=\left( \pm 4,\pm 4\right) $, are $s=-1$.}
\label{fig7}
\end{figure}

The model also supports more complex stable vortex structures, such as
supervortices, see an example of the oblique type in Fig. \ref{fig7}.
Although each compact crater-shaped vortex, of which the structure is built,
is unstable in isolation, the ring formed by four of them, with the global
vorticity imprinted onto it, is \emph{stable}, as verified by direct
simulations. $Q(k)$ curves for families of the supervortices are similar to
those shown in Fig. \ref{fig6}, their stability also precisely obeying the
VK criterion. In particular, for the same values of parameters as in Fig. %
\ref{fig7}, the stability border (which coincides with the point where $qQ/dt
$ vanishes) is found at $k\approx 12$. At this point, the norm of the
supervortex attains its maximum, $Q_{\max}\approx 27$.

\begin{figure}[tbp]
\begin{center}
\subfigure[]{\includegraphics[width=3.6in]{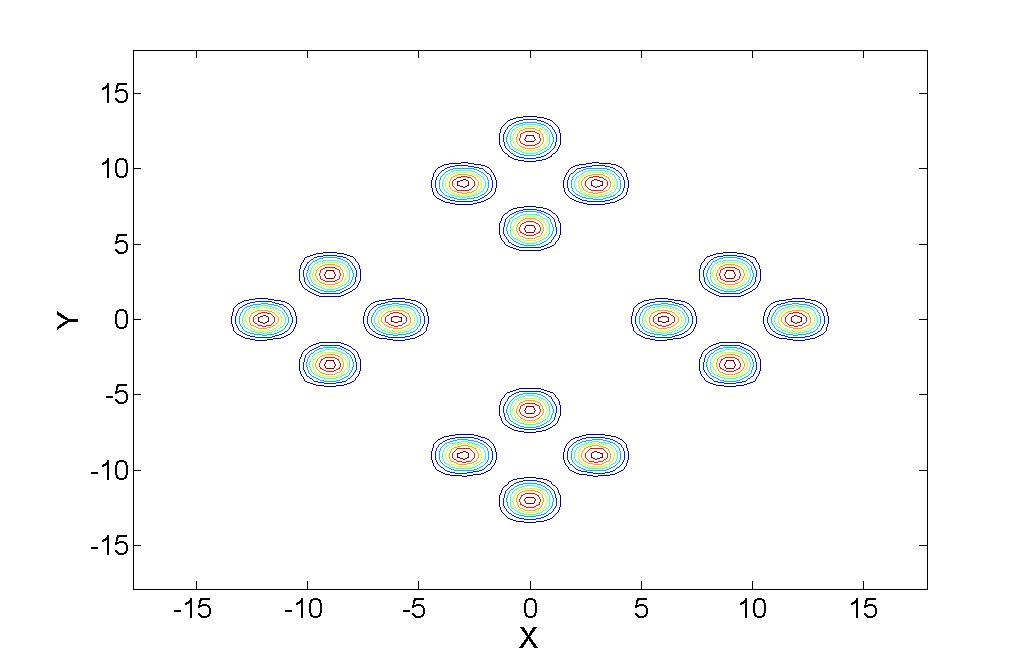}}\subfigure[]{%
\includegraphics[width=3.6in]{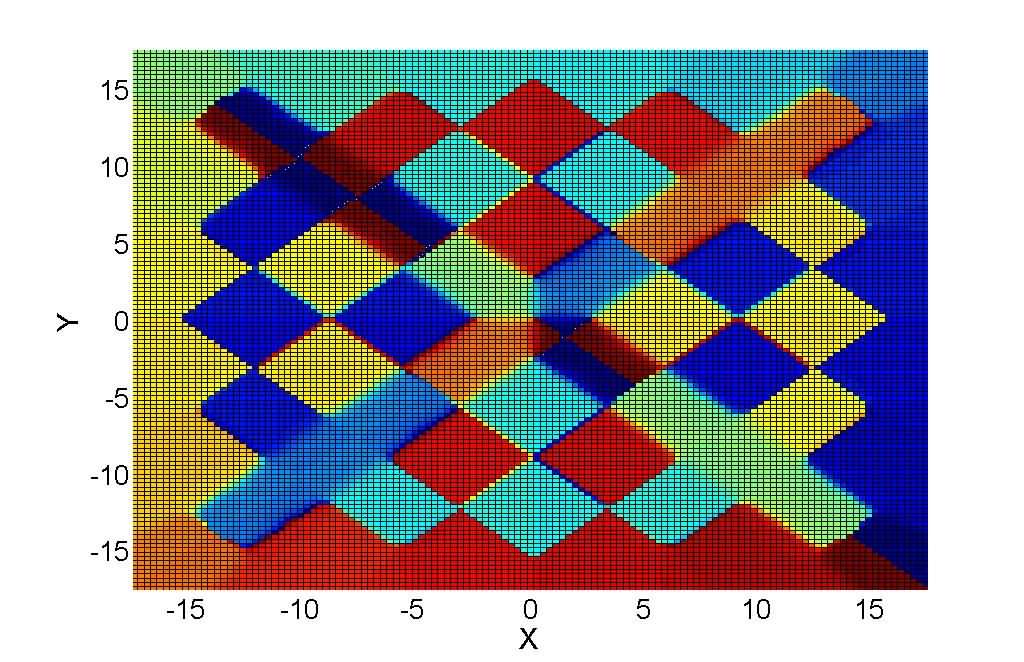}}
\end{center}
\caption{(Color online) An example of a stable \textit{quadrupole vortex}
(of the straight type), built as a chain of four quadrupoles, with the
global phase circulation of $2\protect\pi $. Panels (a) and (b) display the
distribution of the local density and phase in the pattern. Parameters are $%
V_{0}=10$, $D=3$, $\protect\gamma =1$, and $k=8.5$, $Q=1.17$.}
\label{fig8}
\end{figure}

Families of higher-order vortices with $S>1$ have been found too. A novel
type of stable solitary vortices can be constructed using, as building
blocks, a set of four \textit{quadrupoles} (rather than simple peaks, cf.
Fig. \ref{fig5}), with the phase shift of $\pi /2$ between adjacent ones,
which corresponds to $S=1$. An example of a corresponding \textit{quadrupole
vortex} of the straight type is displayed in Fig. \ref{fig8}. These findings
will be reported in a detailed form elsewhere.

\section{Conclusion}

We have demonstrated that various species of solitons and vortices
can be stabilized by periodic potentials in the 2D geometry against
the supercritical collapse, driven by the self-attractive
cubic-quintic nonlinearity. For the Kronig-Penney and $\cos $
potentials, soliton families were obtained by means of the
variational approximation and in a numerical form. The stability of
all the families, including fundamental and multi-humped solitons,
solitary vortices of the oblique and straight types and
supervortices precisely obeys the VK criterion. A novel species of
quadrupole vortices has been demonstrated. The model can be realized
in composite (colloidal) optical media, and in self-attractive BEC
in ``pancake"-shaped traps.

\end{document}